%% file: v2-paper.tex
\documentclass[11pt,a4paper]{article}
\pdfoutput=1  % arXiv: force pdflatex (submission embeds a PDF graphic, moltrust-logo.pdf)

\usepackage[utf8]{inputenc}
\usepackage[T1]{fontenc}
\usepackage{helvet}

\usepackage[margin=1in]{geometry}
\usepackage{setspace}
\usepackage{parskip}

\usepackage{amsmath}
\usepackage{amssymb}
\usepackage{amsthm}

\usepackage{booktabs}
\usepackage{tabularx}
\usepackage{array}
\usepackage{multirow}
\usepackage{longtable}

\usepackage{listings}
\usepackage{xcolor}
\definecolor{codebg}{RGB}{248,248,248}
\definecolor{codestring}{RGB}{163,21,21}
\definecolor{codecomment}{RGB}{100,100,100}
\usepackage{graphicx}
\usepackage{float}
\usepackage{tikz}
\usetikzlibrary{positioning,arrows.meta,shapes.geometric,fit,calc,backgrounds}
\usepackage{standalone}
\input{figures/colors.tex} % shared muted figure palette (single source of colour)

\usepackage{xurl}
\usepackage[colorlinks=true,
            linkcolor=blue!50!black,
            citecolor=blue!50!black,
            urlcolor=blue!50!black,
            pdfauthor={Lars Kersten Kroehl},
            pdftitle={Trust Without Trusting: A Recomputable Trust Protocol for Autonomous Agents},
            pdfsubject={MolTrust Protocol arXiv preprint v2},
            pdfkeywords={agent authorization, enforcement governance, decentralized governance, oracle problem, honest-verifier data availability, Sybil resistance, W3C DID, verifiable credentials, MolTrust, CEP, AAE}]{hyperref}
\usepackage[capitalize,noabbrev]{cleveref}

\usepackage[backend=biber,
            style=ieee,
            sorting=none,
            maxbibnames=10,
            giveninits=true]{biblatex}
\usepackage{fancyhdr}
\title{\textbf{Trust Without Trusting}\\[0.3em]
\large A Recomputable Trust Protocol for Autonomous Agents\\[0.2em]
\normalsize\normalfont Empirical Evidence from a W3C VC + DID Trust Infrastructure}

\author{Lars Kersten Kroehl\\
  \normalsize MolTrust / CryptoKRI GmbH, Zurich\\
  \normalsize \href{mailto:lars@moltrust.ch}{lars@moltrust.ch} $\cdot$ \href{https://moltrust.ch}{moltrust.ch}}

\date{14 June 2026\\[0.9em]
{\footnotesize This preprint (PDF v2.0) is anchored on Base Layer~2 Mainnet under Tag \texttt{MolTrust/arXiv/v2.0}.\\
Integrity record: \href{https://moltrust.ch/publications/integrity.html}{moltrust.ch/publications/integrity.html}}}

\begin{document}

\sloppy
\emergencystretch=3em

\maketitle

% MolTrust logo, top-left corner of the first page (figures/moltrust-logo.pdf, from the SVG source).
\begin{tikzpicture}[remember picture, overlay]
  \node[anchor=north west, inner sep=0pt]
    at ([xshift=1.0cm, yshift=-0.85cm]current page.north west)
    {\includegraphics[height=1.1cm]{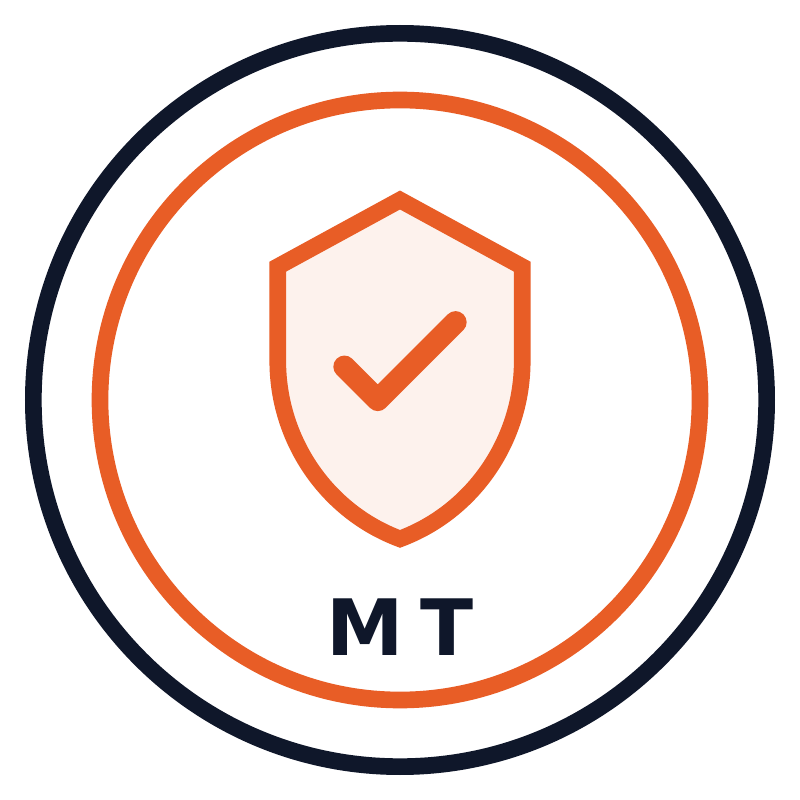}};
\end{tikzpicture}

\input{front-sections.tex}   % Abstract + §1 Introduction + §2 Related Work + §3 Background
\input{spine-sections.tex}   % §4 Problem + §5 CEP + §6 Scope & Limits
\input{back-sections.tex}    % §7 Deployment Evidence + §8 Discussion + §9 Conclusion

\printbibliography

\end{document}

%% file: figures/colors.tex
\definecolor{figline}{HTML}{3A4A5A}        % muted slate-blue — outlines + text (standard)
\definecolor{figfill}{HTML}{ECF1F6}        % very light blue-grey — standard node fill
\definecolor{figaccent}{HTML}{2E6E66}      % muted teal — central / key element outline
\definecolor{figaccentfill}{HTML}{DBEAE6}  % light teal — central element fill
\definecolor{figmuted}{HTML}{6B7785}       % muted grey — labels / secondary text

%% file: front-sections.tex
% =====================================================================
% arXiv v2 — FRONT: Abstract + §1 Intro + §2 Related Work + §3 Background
% TONE PASS 2026-06-12 (Lars): claim-first, POSITIVE. CEP = "Combined Evidence
% Protocol" (acronym kept; E: Enforcement -> Evidence). No "advisory" wording:
% the reference layer "records and signs verdicts any party recomputes".
% Honest limits KEPT, restated positively. Substance/citations/labels unchanged.
% NB: Abstract below is a DRAFT-TO-SPEC — Lars's exact draft was not in the
% message; swap his wording in when available.
% =====================================================================

\begin{abstract}
Autonomous AI agents already transact at production scale — 69{,}000 bots, 165 million transactions, \$50 million in volume on a single marketplace \cite{ref1_agentmarket} — and any party can verify a signed credential without a central service. In an open agent world that covers most of what trust requires: there are no universal borders, and each party chooses for itself whom to deal with.

Borders appear only where a closed space draws one — a marketplace, a platform, or a consortium sets house rules. Whoever draws the border holds the authority to apply it, and may apply it as they choose, behind closed doors. This paper addresses the gap that opens there: when you rely on someone else's border, how do you check that they applied their own published rules — on your own, taking no one's word for it, and handing the check to no new trusted party?

Our answer is the Combined Evidence Protocol (CEP): a five-condition predicate any party recomputes from anchored data, turning ``did the boundary-owner follow its own admission rules'' into a fact anyone verifies rather than a claim anyone believes. The move that secures optimistic rollups secures this — correctness rests on the ability to recompute, so the measurement belongs to everyone and the oracle problem such checks usually carry dissolves. Keyed commitments hold permanent verifiability and the right to erasure together; cluster diversity is proven from graph structure, so fabricated identities yield no fabricated consensus.

CEP earns its place where it matters most: a consortium of co-equal, mutually distrusting peers under a shared charter, each able to verify — independently, continuously — that the rules they jointly agreed are the rules being applied. It belongs to the family of trustless systems — optimistic and zero-knowledge rollups, verifiable ML, self-sovereign-identity predicates — a recomputable claim under mutual mistrust. The infrastructure beneath it is live: a W3C VC + DID trust layer running since March 2026, anchored on Base L2, continuing the deployment of \emph{From Specification to Deployment} \cite{ref57_kroehl_v19} and standing on its own.
\end{abstract}

% ---------------------------------------------------------------------
\section{Introduction}
\label{sec:intro}
% Claim-first, positive. v1.9 continuity. CEP = Combined Evidence Protocol.
% ---------------------------------------------------------------------

Trust between autonomous agents can be established without trusting a central party — by making the exercise of authority \emph{recomputable}: auditable by anyone, while control stays with whoever holds it. The aim is accountability, not a transfer of authority. This paper continues \emph{From Specification to Deployment} \cite{ref57_kroehl_v19}, which documented the deployment of a W3C VC + DID trust infrastructure for autonomous agents; here we develop, in a self-contained way, the governance question that infrastructure raises. The deployment of autonomous AI agents has moved from proof-of-concept into production commerce. A single marketplace reports 69{,}000 autonomous bots executing 165 million transactions across \$50 million USDC in cumulative volume, without any shared trust layer between participants \cite{ref1_agentmarket}. Across financial services, non-human identities now outnumber human employees by wide margins and are growing fast \cite{ref2_neville,ref3_entrolabs}; network data covering a large share of global web traffic shows AI-agent traffic ratios orders of magnitude above human traffic \cite{ref4_prince,ref5_cloudflare}. These are present conditions, not projections.

That infrastructure removes single points of failure for \emph{verification}: any party can check a signed credential without a central service \cite{ref6_w3cdid,ref7_w3cvc}, and a growing body of work shows how to assemble such primitives into agent trust systems (Section~\ref{sec:related}). In an open agent world, verification is most of what is needed — there is no universal boundary to enforce, and no party is asking for a universal enforcer. Boundaries appear only where a \emph{closed space} draws one: a marketplace, a platform, or a consortium exercises house rules; a jurisdiction draws a political border. These boundaries are local and chosen, and whoever runs the closed space can already enforce them. The governance question is different: wherever a boundary is drawn, authority over it accrues to its drawer and can be exercised opaquely. \textbf{How can the exercise of boundary-authority be held to account by parties that distrust its holder, with that accountability resting on recomputation rather than a new trusted party?}

Our answer is \emph{recomputable accountability}. We separate two scopes, because they have different bases of authority:

\begin{itemize}
  \item \textbf{Operator-local enforcement.} A relying party may gate \emph{its own} agent against \emph{its own} authorization, under its own authority — exactly as any deployer bounds its own software. This needs no convention; it is buildable on the deployed system today.
  \item \textbf{A network-wide accountability convention.} An \emph{optional} default that participating operators voluntarily follow, useful when those who rely on a boundary want to verify — without trusting its owner — that the owner applied its own published admission rules. This is the hard scope, and the one CEP addresses.
\end{itemize}

We answer it with the \textbf{Combined Evidence Protocol (CEP)}, which we develop from the anchored specification's ``Combined Enforcement Protocol'' to reflect its evidentiary function (Section~\ref{sec:cep}). CEP supplies a publicly \textbf{recomputable} signal — five conditions about the relying-party population combined into one predicate any party recomputes from anchored data (Section~\ref{sec:cep}) — that holds the exercise of boundary-authority to account. The reference layer records and signs verdicts that any party recomputes; following the convention is voluntary; the transaction decision always rests with the agent or its operator. CEP makes the \emph{use} of boundary-authority auditable; the authority itself stays with whoever holds it — accountability, not control. The pattern CEP belongs to is that of trustless systems — optimistic and zero-knowledge rollups, verifiable ML, self-sovereign-identity predicates: a recomputable claim under mutual mistrust. Its load-bearing setting is a consortium of co-equal, mutually distrusting peers under a shared charter, verifying the admission rules they jointly agreed.

Our contributions are: (1) a statement of the boundary-accountability problem and its oracle sub-problem — how the exercise of authority over a voluntarily-drawn boundary is recomputably audited by recomputation alone (Section~\ref{sec:problem}); (2) the Combined Evidence Protocol, an optional convention whose accountability rests on a conditions-based, publicly recomputable signal that removes the need for a trusted measurement party at the protocol level — the oracle problem (Section~\ref{sec:cep}); (3) an explicit scope-and-limits account, including the operator-local vs.\ network-wide distinction and an honest status (Section~\ref{sec:scope}); and (4) deployment evidence that the system is real — live since March 2026 across eight verticals, with on-chain anchoring (Section~\ref{sec:evidence}).

% ---------------------------------------------------------------------
\section{Related Work and Positioning}
\label{sec:related}
% Material: v1.9 §2 CONDENSED + extended with the three lineage anchors.
% Western anchors mandatory; never an exclusively non-western source base.
% ---------------------------------------------------------------------

\subsection{Regulatory and institutional context}

The Singapore IMDA \textit{Model AI Governance Framework for Agentic AI} (January 2026) is the first comprehensive governance framework for autonomous agents; it requires each agent to carry a verifiable digital identity and an audit trail of which agent acted under whose authorisation \cite{ref8_imda,ref21_imdafull}. NIST's Center for AI Standards and Innovation launched an AI Agent Standards Initiative (February 2026), and the associated NCCoE concept paper frames the gap directly: agents are commonly treated as generic service accounts without dedicated identity, authorization, or accountability controls \cite{ref9_nistcaisi,ref10_nccoe,ref22_cosais}. The EU AI Act (Regulation (EU) 2024/1689) applies Article~14 (Human Oversight) and Article~15 (Accuracy, Robustness, Cybersecurity) to autonomous agents in high-risk domains \cite{ref23_euaiact}; the Article~14 duty falls on the relying party, not the protocol — the distinction Section~\ref{subsec:protocol-not-deployer} sets out.

\subsection{Industry framework convergence}

Anthropic's \textit{Trustworthy Agents in Practice} (April 2026) states that the model layer alone cannot secure agentic AI and calls for shared infrastructure no single vendor can provide, favouring open standards over proprietary alternatives \cite{ref11_anthropic}; Anthropic reaffirmed and operationalised this position in \textit{Zero Trust for AI Agents} (May 2026) \cite{ref58_anthropic_zerotrust}. Google's SAIF~2.0 Agent Risk Map identifies rogue-action and over-permissioned-tool risks and the principles of well-defined controllers, limited powers, and observable actions \cite{ref12_google,ref24_cosai}; Microsoft's Entra Agent ID addresses agent identity within a single enterprise perimeter \cite{ref25_entra}. These efforts establish the need that motivates this work; we build on them and extend further — holding the exercise of authority over a network-wide boundary to account, across organisational boundaries, by recomputation rather than a central authority.

\subsection{Academic foundations}

Closest to this work are deployed-or-prototyped agent trust systems. SAGA \cite{ref16_syros} provides centralised, formally analysed policy enforcement through a trusted Provider — strong within one administrative domain, but it presupposes the very trusted party we seek to avoid at the network scope. Hu and Rong \cite{ref15_hurong} give a six-category taxonomy of inter-agent trust (Brief, Claim, Proof, Stake, Reputation, Constraint); Schroeder de Witt \cite{ref14_schroederdewitt} frames multi-agent security as a field with threats specific to agent interaction \cite{ref26_motwani}. Acharya's TIVA \cite{ref17_acharya} and Rodriguez Garzon et al.\ \cite{ref36_rodriguezgarzon} describe DID + VC + on-chain-anchoring architectures conceptually; Rodriguez Garzon et al.\ explicitly identify the limitation of an LLM in sole charge of security procedures, motivating trust mechanisms that operate independently of the model layer. Threat taxonomies by Ferrag et al.\ \cite{ref18_ferrag} (cross-mapped to CVE/NVD) and Mao et al.\ \cite{ref20_mao} catalogue the attack surface, alongside the OWASP agentic-security material \cite{ref19_owasp}. We build directly on these architectures and extend them: the \emph{recomputable accountability} of a voluntarily-drawn boundary, specifically in a zero-knowledge / mutual-mistrust setting, becomes the object of design.

\subsection{Governance, data-availability, and Sybil-resistance lineages}
\label{subsec:lineages}

CEP draws three lineages from outside the agent-trust literature, each of which informs a specific mechanism in Section~\ref{sec:cep}.

\textbf{On-chain governance and concentration.} Analyses of DAO voting power show that on-chain governance frequently concentrates in a few actors, with low Nakamoto coefficients and weak participation \cite{ref39_daogov}. This concentration is a reference point CEP's concentration caps (Section~\ref{subsec:params}) are designed to beat, and the comparison for why timelocks and per-actor/per-cluster bounds matter.

\textbf{Optimistic-rollup security (verification \texttt{>} production).} Optimistic rollups establish correctness not by trusting a block producer but by letting any verifier recompute and challenge a published claim \cite{ref38_arbitrum}. CEP's honest-verifier data availability (Section~\ref{subsec:hvda}) borrows this stance directly: the evidence predicate is a deterministic function anyone can recompute from anchored data, so no party holds the oracle.

\textbf{Sybil resistance from graph structure.} Social-graph Sybil defences \cite{ref40_sybilguard} and proof-of-personhood / unique-humanity systems \cite{ref41_brightid} establish that resistance to identity fabrication is better grounded in structural relationships than in self-declared attributes. CEP's cluster-diversity condition (Section~\ref{subsec:clusters}) follows this lineage: independence is computed from the endorsement graph, not from declared categories.

\subsection{Position summary}

This paper builds on the verification-and-identity work above and extends it in one respect: its object is the \emph{recomputable accountability} of the use of boundary-authority — evidence no single party produces or owns — a layer above verification and enforcement, relevant specifically where mutual mistrust makes recomputation the point. That object is what scope (b) and CEP address; the deployed system (Section~\ref{sec:background}) shows the question arises in practice.

% ---------------------------------------------------------------------
\section{Background: The Deployed Verification System}
\label{sec:background}
% Material: v1.9 §3.1–3.4 CONDENSED. Short; feeds spine. Positive tone.
% ---------------------------------------------------------------------

This section condenses the deployed system to just enough to ground the governance argument; full normative definitions are in the Protocol Technical Specification \cite{ref28_techspec}. Throughout, a \emph{relying party} is an operator that depends on a boundary and chooses whether to act on the protocol's evidence — the verifier in W3C Verifiable Credentials terms, and, at the network scope, a participant whose adoption the five conditions measure. The point to carry forward is that the system is deployed and produces signed, recomputable verdicts: it verifies and records, and the transaction decision rests with the relying party — so the open question is how an optional network-wide convention built on it makes the use of boundary-authority recomputably accountable.

\textbf{Four primitives.} Each agent holds a W3C Decentralized Identifier \cite{ref6_w3cdid}, with ownership proven by an Ed25519 key; the reference method is \texttt{did:moltrust}, but any method supporting Ed25519/secp256k1 is conformant. Authorization is expressed as a W3C Verifiable Credential \cite{ref7_w3cvc} supporting attenuating delegation chains any verifier can traverse independently. Behaviour is recorded as Interaction Proof Records — dual-signed, Merkle-anchored on Base L2 — accumulating into a registry-computed Trust Score (Section~\ref{sec:evidence}). All three are expressed in open, portable formats, so evidence issued by one party is verifiable by any other.

\begin{figure}[tb]
\centering
\input{figures/fig-primitives.tex}
\caption{The four primitives: a DID identifies the agent, a Verifiable Credential authorizes it, Interaction Proof Records record its behaviour, and the Trust Score aggregates that record.}
\label{fig:primitives}
\end{figure}
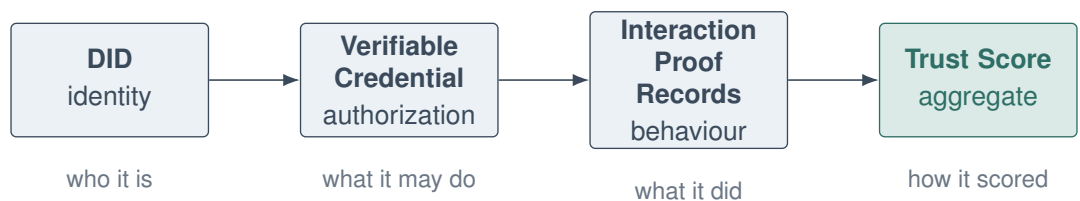

\textbf{The Agent Authorization Envelope (AAE).} The machine-evaluable authorization object carries three normative blocks: \textbf{MANDATE} (permitted purpose, allowed/denied action patterns, delegation rules), \textbf{CONSTRAINTS} (time windows, financial thresholds, jurisdictions, counterparty minimums, obligations), and \textbf{VALIDITY} (issuer, holder binding, mandatory expiry, revocation, optional anchor). Any conformant evaluator applies default-deny, deny-precedence, attenuation-only delegation, and mandatory expiry. The design reuses NIST SP~800-162 ABAC \cite{ref29_abac}, RFC~9396 \cite{ref30_rar}, RFC~8785 JCS \cite{ref31_jcs}, and Ed25519Signature2020 \cite{ref32_ed25519}.

\begin{figure}[tb]
\input{snippets/snippet-aae.tex}
\caption{A representative Agent Authorization Envelope. A commerce agent may place orders but is denied the admin path (deny takes precedence over allow); it is bounded by a per-transaction value, a rate limit, a time window, and a counterparty-trust floor, with mandatory expiry and a revocation endpoint.}
\label{lst:aae}
\end{figure}

\textbf{Three-layer architecture.} Authorization is checked at three layers: \emph{cryptographic} (Ed25519 + RFC~8785, tamper-evident by construction), \emph{API} (trust-score and revocation consequences, fail-closed on unreachable revocation), and \emph{kernel} (Falco eBPF detection below the agent's process boundary \cite{ref13_falcobridge}). In the reference deployment a DENY is produced as a signed, anchorable verdict that any party recomputes, and recorded; whether to act on the verdict is the relying party's decision \cite{ref28_techspec}. Operator-local blocking (scope (a), Section~\ref{subsec:protocol-not-deployer}) is available to a relying party in its own gateway; at the network scope the protocol produces recomputable evidence that operators opt into, rather than a chokepoint imposed on them.

\textbf{Decentralised evidence, centralised registry — for now.} The evidence layer is verifier-independent: any party with an issuer's key validates any signed artifact without a central service. The reference registry, however, is run by one organisation, and how an optional network-wide convention on that infrastructure makes the use of boundary-authority recomputably accountable — without that accountability resting on the single operating organisation — is the subject of the rest of the paper.

\begin{figure}[b]
\centering
\input{figures/fig-three-layer.tex}
\caption{The three-layer architecture: each layer states what it checks and at which boundary, from cryptographic integrity through registry-level consequences to kernel-level detection.}
\label{fig:three-layer}
\end{figure}

%% file: figures/fig-primitives.tex
\begin{tikzpicture}[
  node distance=12mm,
  pbox/.style={draw=figline, line width=0.5pt, rounded corners=2pt, align=center,
               text=figline, font=\small\sffamily, fill=figfill,
               text width=24mm, minimum height=15mm, inner sep=3pt},
  pacc/.style={pbox, draw=figaccent, fill=figaccentfill, text=figaccent},
  flow/.style={-{Latex[length=2.2mm]}, draw=figline, line width=0.6pt},
  clab/.style={font=\footnotesize\sffamily, text=figmuted}
]
\node[pbox] (did) {\textbf{DID}\\[2pt]identity};
\node[pbox, right=of did] (vc)  {\textbf{Verifiable Credential}\\[2pt]authorization};
\node[pbox, right=of vc]  (ipr) {\textbf{Interaction Proof Records}\\[2pt]behaviour};
\node[pacc, right=of ipr] (ts)  {\textbf{Trust Score}\\[2pt]aggregate};
\draw[flow] (did) -- (vc);
\draw[flow] (vc)  -- (ipr);
\draw[flow] (ipr) -- (ts);
\node[clab, below=3mm of did] {who it is};
\node[clab, below=3mm of vc]  {what it may do};
\node[clab, below=3mm of ipr] {what it did};
\node[clab, below=3mm of ts]  {how it scored};
\end{tikzpicture}

%% file: snippets/snippet-aae.tex
% snippet-aae — §3, a realistic Agent Authorization Envelope (MANDATE/CONSTRAINTS/VALIDITY).
% \input-able bare lstlisting (the paper wraps it in a float with caption).
% Reusable as source. Parent loads listings.
\begin{lstlisting}[basicstyle=\ttfamily\footnotesize,breaklines=true,frame=single,
                   framesep=4pt,xleftmargin=2pt,showstringspaces=false]
{
  "mandate": {
    "purpose": "commerce",
    "allow":  ["https://api.shop.example/v1/orders/*"],
    "deny":   ["https://api.shop.example/v1/admin/*"],
    "delegation": { "allowed": false }
  },
  "constraints": {
    "max_transaction_value": { "amount": 250000, "currency": "USDC" },
    "rate_limit":   "PT1H;max=20",
    "allowed_hours": { "tz": "Europe/Zurich", "from": "08:00", "to": "20:00" },
    "counterparty_min_trust": 60
  },
  "validity": {
    "issuer":     "did:moltrust:bank-9f2a",
    "holder":     "did:moltrust:agent-7c3a",
    "not_before": "2026-06-01T00:00:00Z",
    "not_after":  "2026-09-01T00:00:00Z",
    "revocation": "https://api.moltrust.ch/revocation/aae/7c3a"
  }
}
\end{lstlisting}

%% file: figures/fig-three-layer.tex
\begin{tikzpicture}[
  lyr/.style={draw=figline, line width=0.4pt, align=left, text=figline, font=\small\sffamily,
              fill=figfill, text width=88mm, minimum height=16mm, inner sep=6pt, anchor=north}
]
% three layers stacked top-to-bottom, sharing borders (a true stack)
\node[lyr] (l1) at (0,0)
  {\textbf{Layer 1 \ Cryptographic}\\[2pt]\footnotesize Ed25519 over RFC~8785 canonical JSON: AAE boundaries tamper-evident, verifiable without infrastructure.};
\node[lyr, below=0pt of l1.south, anchor=north] (l2)
  {\textbf{Layer 2 \ API}\\[2pt]\footnotesize Registry-level trust-score and revocation consequences; fail-closed on an unreachable revocation endpoint.};
\node[lyr, below=0pt of l2.south, anchor=north] (l3)
  {\textbf{Layer 3 \ Kernel}\\[2pt]\footnotesize Falco eBPF detection below the agent's process boundary, recording violations the agent runtime cannot suppress.};
% enclosing architecture frame (accent)
\node[draw=figaccent, line width=0.9pt, rounded corners=2pt, fit=(l1)(l2)(l3), inner sep=2.5pt] (frame) {};
\end{tikzpicture}

%% file: spine-sections.tex
% =====================================================================
% arXiv v2 — SPINE: §4 problem, §5 CEP, §6 scope
% TONE PASS 2026-06-12 (Lars): claim-first, POSITIVE. CEP = "Combined Evidence
% Protocol" (acronym kept). No "advisory": the layer "records and signs
% verdicts any party recomputes". The 5-AND combines five conditions into
% RECOMPUTABLE EVIDENCE about the boundary. Honest limits KEPT, restated
% positively (esp. §6.2). Substance/citations/labels/parameters unchanged.
% Source: TECH_SPEC v0.9 §17, anchored Base L2 block 46,986,137 (ref28).
% =====================================================================

% ---------------------------------------------------------------------
\section{The Boundary-Accountability Problem}
\label{sec:problem}
% Material: TECH_SPEC §17.2.1 + ADR CEP-Governance + v1.9 §3.5 (promoted).
% ---------------------------------------------------------------------

Wherever a boundary is drawn, authority over it accrues to its drawer — the power to decide who is admitted and who is excluded — and that authority can be exercised opaquely. Section~\ref{sec:background} described the MolTrust layer as it runs today: it verifies an Agent Authorization Envelope (AAE), evaluates a proposed action against the envelope's MANDATE, CONSTRAINTS, and VALIDITY blocks, and emits a signed, anchorable verdict that any party recomputes; whether to act on it is the relying party's decision \cite{ref28_techspec}. An operator can already act on a verdict for its \emph{own} agent (Section~\ref{subsec:protocol-not-deployer}). The harder question is network-wide, and it is the question this paper sets out to answer:

\begin{quote}
\emph{How can the use of boundary-authority be held to account by parties that distrust its holder?}
\end{quote}

\noindent The aim is accountability that is itself trustworthy — evidence anyone can recompute — resting on recomputation rather than a new trusted party.

This is what the Combined Evidence Protocol provides, and it provides it \emph{as evidence}: the protocol audits how authority is used, while the authority itself stays with whoever holds it (Section~\ref{sec:scope}). The rest of this section states why the natural ways to certify such evidence fail, and the oracle problem any conditions-based scheme must dissolve.

\subsection{Why the obvious certifiers fail}

The party that certifies the evidence — that says, network-wide, ``these conditions about the relying-party population hold'' — must be one no one has to trust, and it must remain so over the infrastructure's intended lifetime, measured in years. Three intuitive certifiers each fail under that horizon \cite{ref28_techspec}:

\begin{itemize}
  \item \textbf{A person.} Binding the judgement to a founder or a named operator lasts only as long as that person: people leave, organisations are acquired, keys are lost or compelled. Evidence only one person can vouch for is mere testimony.
  \item \textbf{A single chain.} Anchoring the judgement to one ledger inherits that ledger's mortality and politics: it can halt, censor, reorganise, or fork. A verification protocol meant to be neutral keeps its most consequential signal free of any single chain's governance.
  \item \textbf{A single instance.} Vesting the judgement in one running deployment recreates a single point of failure of exactly the kind that motivated decentralised identity in the first place: the instance can be shut down, seized, or compromised, and with it the signal.
\end{itemize}

The design replaces all three with \textbf{objective, publicly recomputable conditions}: the evidence holds because a set of measurable facts about the relying-party population holds — a fact anyone can check — not because any party certified it.

\subsection{The oracle problem}
\label{subsec:oracle}

Conditions-based evidence moves the difficulty rather than removing it. If the evidence reads ``enough independent relying parties have adopted the boundary,'' then \emph{someone} must measure adoption — count the relying parties, judge their independence, decide the threshold is met. Whoever holds that measurement becomes a new trusted party: an oracle. An oracle that can misreport the world produces or withholds the evidence at will, which reintroduces precisely the single point of authority the conditions were meant to eliminate.

This is the pivot on which the paper turns. We name it the \textbf{oracle problem for recomputable evidence}: \emph{a conditions-based signal is only as trustworthy as the party that measures whether the conditions are met}. Recomputable accountability is worth little if a single measurer can decide what the data says — that party would be the central authority under a new name, and a scheme that leaves the problem unaddressed merely relabels its operator a ``measurement service.'' Section~\ref{sec:cep} sets out the Combined Evidence Protocol (CEP), whose central move (Section~\ref{subsec:hvda}) makes the measurement a deterministic function anyone can recompute, so that no party holds the oracle.

% ---------------------------------------------------------------------
\section{The Combined Evidence Protocol (CEP)}
\label{sec:cep}
% Core contribution. Material: TECH_SPEC v0.9 §17.2 (PRIMARY, anchored
% block 46,986,137 = ref28_techspec) + ADR CEP-Governance + CEP-3
% Threshold Spec. Summarise + reference; do NOT re-derive the design.
% ---------------------------------------------------------------------

CEP is the model for the recomputable evidence introduced in Section~\ref{sec:problem}. Its full design is fixed in the MolTrust \emph{CEP Governance} architecture decision record and the \emph{CEP-3 Threshold Specification}, and is published at roadmap level — with parameters and rationale — in \S17.2 of the Protocol Technical Specification v0.9, which is anchored on Base L2 as an immutable, independently citable record of the design as of its acceptance.\footnote{TECH\_SPEC v0.9 \S17, anchored at Base L2 block 46{,}986{,}137 under payload \texttt{MolTrust/TechSpec/0.9} (sha256:\texttt{462af65a\ldots}); see \cite{ref28_techspec}. The specification names the protocol ``Combined Enforcement Protocol''; we forward-develop it here as the \emph{Combined Evidence Protocol}, reflecting that its function is recomputable evidence rather than enforcement. We summarise the design here so the paper is self-contained, and treat the anchored specification, not this paper, as the authoritative source.} The purpose of this section is to make the mechanism legible, not to re-derive it: each subsection states what the mechanism does and why it answers a specific failure from Section~\ref{sec:problem}.

\subsection{Five conditions, combined (AND)}
\label{subsec:ramp}

CEP combines five conditions about the relying-party population into a single predicate; the evidence holds only when all five hold \emph{simultaneously}, and acting on it stays the relying party's choice \cite{ref28_techspec}:

\begin{enumerate}
  \item a minimum elapsed time since the thresholds were fixed and anchored (a timelock and public-veto window);
  \item a minimum number $N$ of Sybil-qualified relying parties;
  \item distribution of those relying parties over a minimum number $K$ of independent clusters;
  \item no single actor holding more than a share $X$ of the voting weight;
  \item no single cluster holding more than a share $Y$ of the total.
\end{enumerate}

The conjunction is the point — and the source of the name. Any single threshold — ``$N$ relying parties'' alone — is manipulable: an adversary mints identities until the count is met. \emph{Combining} all five means an adversary must simultaneously clear a Sybil filter, populate multiple independent clusters, and stay under both a per-actor and a per-cluster cap, while a timelock exposes the attempt to public veto. The conditions form a conjunction of gates that must all open, and the cost of opening them jointly is what makes the combined predicate strong evidence rather than a single gameable number. The conjunction of all five — their logical AND — \emph{is} the CEP signal: the recomputable evidence holds only when all five conditions hold at once, and a single missing condition yields no signal.

\begin{figure}[tb]
\centering
\input{figures/fig-cep-conditions.tex}
\caption{The five conditions combined. Evidence holds only when all five gates open at once; any one missing yields no signal. The conjunction, not any single threshold, is the load-bearing element.}
\label{fig:cep-conditions}
\end{figure}
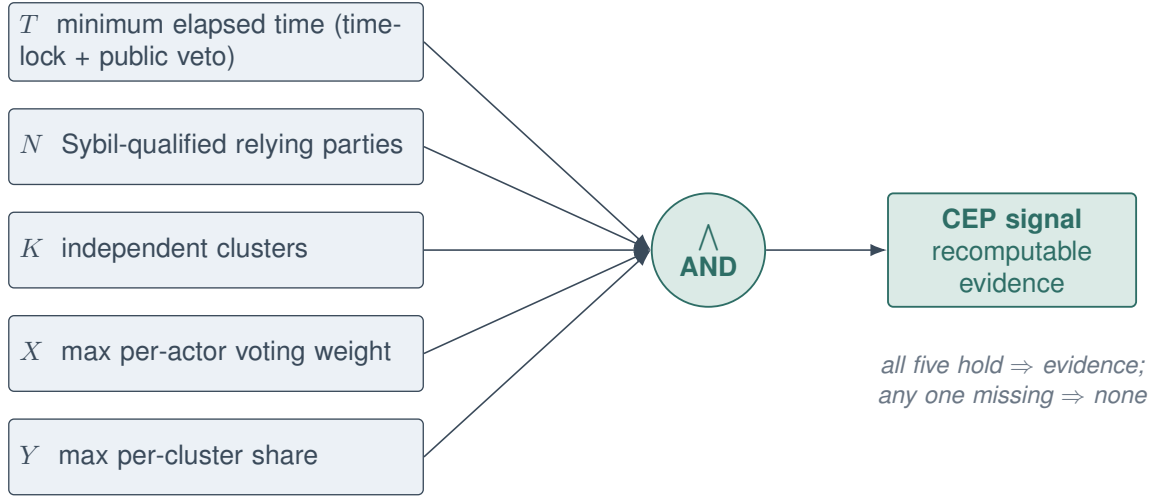

\subsection{Honest-verifier data availability: dissolving the oracle}
\label{subsec:hvda}

The five conditions are facts about the relying-party population, and Section~\ref{subsec:oracle} warned that whoever measures those facts becomes an oracle. CEP removes the oracle rather than trusting it. The condition data is published to decentralised, permanent storage and anchored as a \texttt{(merkle\_root, data\_uri)} tuple across multiple chains; the evidence is then a \textbf{deterministic function that any party can recompute} from the published data \cite{ref28_techspec}. There is no privileged measuring party: a relying party, a skeptic, or a regulator can independently fetch the committed data, recompute the five-condition predicate, and arrive at the same answer.

\begin{figure}[tb]
\centering
\input{figures/fig-recompute-flow.tex}
\caption{Verification $>$ production. Any verifier fetches the committed $(\textsf{merkle\_root},\,\textsf{data\_uri})$, recomputes the five-condition predicate, and reaches the same result; there is no privileged measurer.}
\label{fig:recompute-flow}
\end{figure}
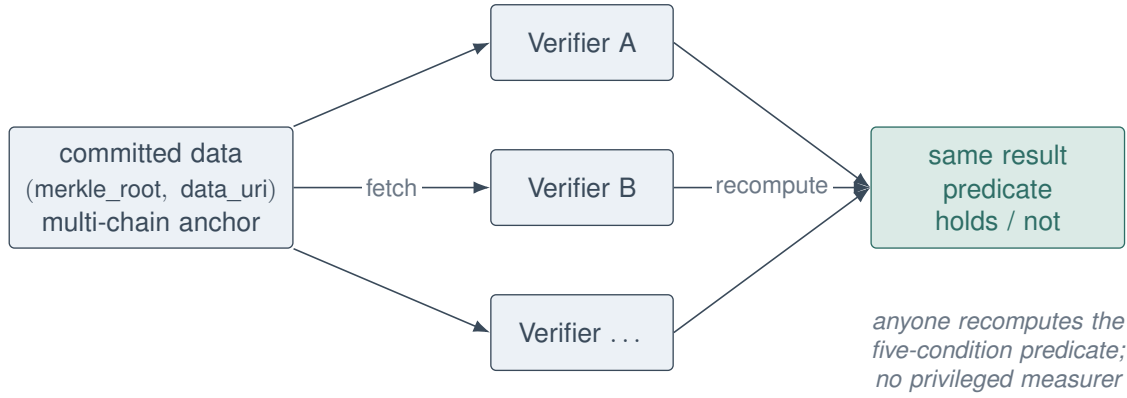

We call this stance \emph{verification $>$ production}: the design lets anyone verify that the conditions were met rather than asking them to trust it, making the determination cheap to verify and impossible to monopolise. The lineage is the same security argument that underwrites optimistic rollups, where correctness rests not on trusting a producer but on the ability of any verifier to recompute and challenge a published claim \cite{ref38_arbitrum}.
Multi-chain anchoring of the commitment tuple means the determination also survives the failure or censorship of any one chain, addressing the single-chain certifier of Section~\ref{sec:problem} at the data layer as well.

\begin{figure}[tb]
\input{snippets/snippet-recompute.tex}
\caption{The recompute predicate. A verifier reads the anchored commitment, downloads and checks the committed data, rebuilds the qualified set and clusters, and evaluates the five conditions; their conjunction is the evidence. Here $\textsf{weight}(\cdot)$ is the governance trust-weight fixed by the CEP-3 threshold specification (Section~\ref{subsec:params}) and $\textsf{sybil\_clear}$ is the registry's Sybil filter (Section~\ref{subsec:sybil}); every input is read from the committed, anchored data, so the result is recomputable from public data rather than a registry judgement.}
\label{lst:recompute}
\end{figure}

\subsection{Keyed commitment and cryptographic erasure}
\label{subsec:erasure}

Publishing condition data permanently collides with data-protection law: relying parties are identifiable entities, and the right to erasure (e.g.\ Art.~17 GDPR) is difficult to reconcile with an immutable anchor. CEP reconciles the two by anchoring only \textbf{keyed commitments} to relying-party identifiers, while the cleartext lives in the permissioned data layer. Erasure is performed by destroying the key (\textbf{cryptographic erasure}), after which the anchored commitment is non-attributable — the integrity proof remains, but it can no longer be tied to a person \cite{ref28_techspec}. Permanent verifiability and the right to be forgotten are thus held at once, rather than traded off.

\subsection{Staged verification: permissioned ramp-up to zero-knowledge}
\label{subsec:staged}

The recomputability of Section~\ref{subsec:hvda} is introduced in two stages. A \textbf{permissioned ramp-up phase} comes first: verifiers are bound by data-processing agreements and recompute the predicate over the committed data under contract. A \textbf{target phase} then replaces that contractual trust with \textbf{zero-knowledge verification}, which proves the recomputation was performed correctly \emph{without disclosing} the underlying relying-party data \cite{ref28_techspec}. The staging is honest about status: the mechanism is exercised first under permissioned assumptions, and the privacy-preserving end-state is a later, separately built phase (Section~\ref{subsec:open}). The zero-knowledge component strengthens the \emph{verification} of the conditions; it is evidence machinery, not authority over agents.

\subsection{Cluster diversity proven from structure}
\label{subsec:clusters}

Condition~3 — distribution over $K$ independent clusters — is the design's defence against an adversary who satisfies a headcount with sockpuppets. Independence is therefore computed from \textbf{structure}, not from \emph{declared} categories (a relying party asserting ``I am a different kind of organisation''), which an adversary controls for free. Clusters are derived from the relying-party endorsement graph by reciprocal-Jaccard analysis — the same construction the registry already runs over agent endorser sets for Sybil detection at the trust-scoring layer (Section~\ref{subsec:sybil}), here applied to the relying-party population. Two relying parties whose endorsement neighbourhoods overlap above a reciprocal threshold collapse into one cluster regardless of how they describe themselves, so diversity is earned in the endorsement graph rather than asserted in labels. Diversity is thus a property the public can recompute from the committed graph, consistent with Section~\ref{subsec:hvda}; grounding independence in graph structure rather than declared attributes follows the Sybil-resistance lineage of Section~\ref{subsec:lineages} \cite{ref40_sybilguard,ref41_brightid}.

\subsection{Parameters and gates}
\label{subsec:params}

The thresholds differ by network. The testnet values exist to \emph{demonstrate} the mechanism end to end; the mainnet values are the production \emph{target}. The two sets serve distinct roles \cite{ref28_techspec}.

\begin{table}[ht]
\centering
\small
\begin{tabular}{@{}llcc@{}}
\toprule
 & \textbf{Parameter} & \textbf{Testnet (demonstration)} & \textbf{Mainnet (target)} \\
\midrule
$N$ & Sybil-qualified relying parties & 11 & 101 \\
$K$ & Independent clusters & 3 & 4 \\
$Y$ & Max share per cluster & $\ge 34\%$ & $33\%$ \\
$X$ & Max voting weight per actor & $10\%$ & $10\%$ \\
$T$ & Timelock / public-veto window & 31 days & 31 days \\
\bottomrule
\end{tabular}
\caption{CEP thresholds. Testnet values demonstrate the mechanism; mainnet values are the production target. Source: \S17.2.3 of the anchored Technical Specification \cite{ref28_techspec}.}
\label{tab:cep-params}
\end{table}

The cluster count differs by design. Informally, by analogy with the byzantine-tolerance relation $N \ge 3f+1$, the mainnet $K = 4$ leaves headroom for one captured or faulty cluster ($f = 1$) whereas the testnet $K = 3$ corresponds to $f = 0$. We use this as a heuristic for sizing, not a hard fault-tolerance guarantee: structural independence in the endorsement graph is the design's proxy for fault-domain independence, and it is only as strong as that proxy. The invariant $Y \ge 1/K$ is preserved in both rows (testnet $K = 3 \Rightarrow Y \ge 34\% > 33.3\%$; mainnet $K = 4 \Rightarrow Y = 33\% \ge 25\%$), so both configurations are structurally satisfiable. These thresholds are also markedly stricter than the concentration limits typical of on-chain DAO governance, where a single actor or a small coalition can frequently meet a quorum \cite{ref39_daogov}.

Two independent gates govern when the combined evidence carries weight, and both are met before it does \cite{ref28_techspec}:

\begin{itemize}
  \item \textbf{Gate-1 — thresholds fixed and anchored.} The parameters of Table~\ref{tab:cep-params} are written down and chain-agnostically anchored before the ramp-up begins. \emph{Status: the values are fixed (CEP-3 Threshold Specification); chain-agnostic anchoring is pending the multi-chain prerequisite.}
  \item \textbf{Gate-2 — prerequisites built.} A relying-party registry with cryptographic (DID-bound) identity and cluster attribution; an explicit \emph{convention-status} signal in the authorization model; and multi-chain anchoring. \emph{Status: open.}
\end{itemize}

Until both gates are met, the layer records and signs verdicts that any party recomputes; the combined evidence carrying network-wide weight is a deliberate, separate step — not a side effect of publishing this paper or the specification it summarises.

% ---------------------------------------------------------------------
\section{Scope and Honest Limits}
\label{sec:scope}
% Material: TECH_SPEC §17.2.2 (scope position) + ADR CEP-Governance + v1.9 §7.
% Negative-lists restated POSITIVELY; honest limits kept.
% ---------------------------------------------------------------------

A design that produces consequential evidence about how a network treats agents invites a fair question about responsibility and overreach. This section states the protocol's scope, the limits of what we claim, and the problems that remain open — before the deployment evidence of Section~\ref{sec:evidence}, so that the evidence is read against an honest boundary.

\subsection{Operator-local enforcement, a network-wide accountability convention, and a protocol layer that is not a deployer}
\label{subsec:protocol-not-deployer}

A DENY verdict (Section~\ref{sec:background}) becomes a \emph{block} only when some party refuses the action it describes. \emph{Who} that party is, and under \emph{whose authority} it acts, separates two scopes this paper is careful to keep apart.

\textbf{(a) Operator-local enforcement.} A relying party may run a fail-closed chokepoint in its \emph{own} gateway: it submits its own agent's proposed action to the evaluator and refuses the action on a signed DENY against its own AAE mandate, under its own authority and with its own override. This scope requires \textbf{no convention} — the operator is already the authorising party for its own agent, exactly as any deployer may bound its own software. In the reference implementation it is buildable today: the registry-side verdict service is deployed (signed, default-DENY verdicts scoped to a mandate and an agent identity), and what remains is the thin caller-side wiring in the operator's own gateway. The operator that gates its own agent also carries the corresponding duty of human oversight.

\textbf{(b) A network-wide accountability convention.} The harder scope is the recomputable evidence of Section~\ref{sec:cep}: an \emph{optional} convention that participating operators voluntarily follow, by which anyone who relies on a boundary can independently verify that the owner applied its own published admission rules. Operators who follow it gain shared, recomputable evidence; others transact as they choose. CEP makes the \emph{use} of boundary-authority auditable while leaving the authority itself with whoever holds it — accountability, which complements control rather than claiming it.

The verdict is identical in both scopes; they differ only in \textbf{who acts on it and under whose authority}. Scope (a) needs no convention; scope (b) is the voluntary network coordination layered above it. What the paper makes recomputable is the \emph{evidence} of how boundary-authority was exercised — a fact anyone can check — and it does so without taking authority over anyone's agent.

This is why MolTrust is a \textbf{verification and evidence \emph{protocol} layer, not the deployer} of a high-risk AI system. The analogy is TLS and the public-key infrastructure: a certificate authority issues certificates and lets any party verify them, while each session and the systems that rely on them remain the relying parties' own to operate \cite{ref28_techspec}. Under both scopes the duty of \emph{human oversight} over a high-risk deployment — including under Art.~14 of the EU AI Act \cite{ref23_euaiact} — rests with the \textbf{relying party} that uses the protocol to gate its agents. The protocol provides oversight-\textbf{enabling} machinery: signed and independently recomputable verdicts, a default-DENY evaluation rule, an append-only audit trail, and a public-veto window during the ramp. It equips a relying party to discharge its duty, which remains the relying party's own to discharge.

\subsection{Status, and what ``demonstrable'' means}
\label{subsec:designed-not-activated}

We are precise about status, and we state it as what the system does. In the reference deployment the layer records and signs verdicts that any party recomputes; operator-local blocking (scope (a) of Section~\ref{subsec:protocol-not-deployer}) is available to a relying party in its own gateway. The combined network-wide evidence is designed and its parameters anchored; the gates of Section~\ref{subsec:params} carry it to network-wide weight. The mechanism is \textbf{demonstrable on testnet}: the testnet parameters of Table~\ref{tab:cep-params} let the five-condition predicate, the recomputable evidence, and the cluster-diversity check run end to end over committed data, which establishes that the mechanism works. What runs is the five-condition \emph{computation}; the Gate-2 production prerequisites of Section~\ref{subsec:params} — a DID-bound relying-party registry with cluster attribution, and multi-chain anchoring — remain open.

Three claims fix the standard precisely. The mechanism produces \textbf{recomputable evidence}, and the transaction decision rests with the relying party. The combined evidence is \textbf{designed and gated}, and the mechanism is demonstrable on testnet. And the property we claim is \textbf{verifiability}: the predicate is publicly recomputable and the mechanism demonstrable — a verifiability claim, which is what the design delivers. Where the deployment evidence in Section~\ref{sec:evidence} reports conformance, it holds to the same standard: four of the five AIP/IBCT features hold against an independent verifier, with expressive chained policy tracked as an open item.

\subsection{Open problems}
\label{subsec:open}

Stating precisely what is unfinished is part of the contribution:

\begin{itemize}
  \item \textbf{Treasury governance.} CEP provides recomputable evidence about an \emph{optional} convention; decentralising the funding and management of the infrastructure is a separate problem, and the treasury remains a governance anchor addressed elsewhere.
  \item \textbf{The zero-knowledge phase.} The privacy-preserving end-state of Section~\ref{subsec:staged} — proving recomputation without disclosing relying-party data — is designed; the demonstrable mechanism today runs under the permissioned, data-processing-agreement assumptions of the ramp-up phase.
  \item \textbf{Weak subjectivity for late joiners.} A party joining long after the evidence first carries weight obtains a trustworthy recent reference point to verify the committed history, rather than recomputing from genesis; bounding and distributing that reference point is an open design question.
\end{itemize}

The contribution is a recomputable-evidence design with a demonstrable mechanism and a clearly bounded set of open problems.

%% file: figures/fig-cep-conditions.tex
\begin{tikzpicture}[
  cond/.style={draw=figline, line width=0.5pt, rounded corners=2pt, align=left,
               text=figline, font=\small\sffamily, fill=figfill,
               text width=52mm, minimum height=10mm, inner sep=4pt},
  gate/.style={draw=figaccent, line width=0.7pt, circle, text=figaccent,
               font=\small\sffamily\bfseries, minimum size=15mm, fill=figaccentfill, align=center},
  obox/.style={draw=figaccent, line width=0.7pt, rounded corners=2pt, align=center,
               text=figaccent, font=\small\sffamily, fill=figaccentfill,
               text width=30mm, minimum height=15mm, inner sep=4pt},
  flow/.style={-{Latex[length=2.2mm]}, draw=figline, line width=0.6pt},
  node distance=3.5mm
]
\node[cond] (c1) {$T$\ \ minimum elapsed time (timelock + public veto)};
\node[cond, below=of c1] (c2) {$N$\ \ Sybil-qualified relying parties};
\node[cond, below=of c2] (c3) {$K$\ \ independent clusters};
\node[cond, below=of c3] (c4) {$X$\ \ max per-actor voting weight};
\node[cond, below=of c4] (c5) {$Y$\ \ max per-cluster share};
\node[gate, right=30mm of c3] (and) {$\bigwedge$\\AND};
\node[obox, right=16mm of and] (sig) {\textbf{CEP signal}\\recomputable evidence};
\foreach \c in {c1,c2,c3,c4,c5} \draw[flow] (\c.east) -- (and.west);
\draw[flow] (and) -- (sig);
\node[font=\footnotesize\sffamily\itshape, text=figmuted, below=5mm of sig,
      text width=36mm, align=center]
     {all five hold $\Rightarrow$ evidence;\\ any one missing $\Rightarrow$ none};
\end{tikzpicture}

%% file: figures/fig-recompute-flow.tex
\begin{tikzpicture}[
  dbox/.style={draw=figline, line width=0.5pt, rounded corners=2pt, align=center,
               text=figline, font=\small\sffamily, fill=figfill,
               text width=34mm, minimum height=16mm, inner sep=5pt},
  dacc/.style={dbox, draw=figaccent, fill=figaccentfill, text=figaccent, text width=30mm},
  vrf/.style={draw=figline, line width=0.5pt, rounded corners=2pt, align=center,
              text=figline, font=\small\sffamily, fill=figfill,
              text width=22mm, minimum height=10mm, inner sep=3pt},
  flow/.style={-{Latex[length=2.2mm]}, draw=figline, line width=0.6pt},
  clab/.style={font=\footnotesize\sffamily, text=figmuted},
  node distance=9mm and 26mm
]
\node[dbox] (anchor) {committed data\\[1pt]{\footnotesize $(\textsf{merkle\_root},\ \textsf{data\_uri})$}\\[1pt]multi-chain anchor};
\node[vrf, right=of anchor] (v2) {Verifier B};
\node[vrf, above=of v2] (v1) {Verifier A};
\node[vrf, below=of v2] (v3) {Verifier $\dots$};
\node[dacc, right=of v2] (res) {same result\\[1pt]predicate holds / not};
\draw[flow] (anchor.east) -- node[clab, midway, fill=white, inner sep=1.5pt] {fetch} (v2.west);
\draw[flow] (anchor.north east) -- (v1.west);
\draw[flow] (anchor.south east) -- (v3.west);
\draw[flow] (v2.east) -- node[clab, midway, fill=white, inner sep=1.5pt] {recompute} (res.west);
\draw[flow] (v1.east) -- (res.west);
\draw[flow] (v3.east) -- (res.west);
\node[clab, font=\footnotesize\sffamily\itshape, below=7mm of res,
      text width=46mm, align=center]
     {anyone recomputes the five-condition predicate; no privileged measurer};
\end{tikzpicture}

%% file: snippets/snippet-recompute.tex
% snippet-recompute — §5.2, how any verifier recomputes the five-condition predicate.
% \input-able bare lstlisting (the paper wraps it in a float with caption).
% Pseudocode: fetch -> recompute 5 conditions -> AND -> result. Reusable as source.
% weight(.) and sybil_clear(.) are fixed normatively (see caption) so every input is
% read from the committed, anchored data.
\begin{lstlisting}[basicstyle=\ttfamily\footnotesize,breaklines=true,frame=single,
                   framesep=4pt,xleftmargin=2pt,mathescape=true]
# inputs: the anchored commitment + the fixed thresholds (N, K, X, Y, T, T_min);
#   weight(.)   = governance trust-weight, fixed by the CEP-3 threshold spec (Sec. 5.6)
#   sybil_clear = the registry's three-mechanism Sybil filter (Sec. 7.3)
(merkle_root, data_uri, anchored_at) := read_anchor()      # any chain
data   := download(data_uri); assert merkle(data) == merkle_root
vetoes := { v in data.vetoes : sybil_clear(v.by) }         # qualified public vetoes

Q := { rp in data : rp.active and trust(rp) >= T_min and sybil_clear(rp) }
clusters := connected_components( reciprocal_jaccard(Q, 0.8) )
W := sum_{rp in Q} weight(rp)

c1 := (now - anchored_at) >= T  and  vetoes == {}   # timelock elapsed AND no qualified veto
c2 := |Q|                                  >= N     # Sybil-qualified relying parties
c3 := |clusters|                           >= K     # independent clusters
c4 := max_{rp in Q} weight(rp) / W         <= X     # per-actor cap
c5 := max_{cl in clusters} weight(cl) / W  <= Y     # per-cluster cap

return c1 and c2 and c3 and c4 and c5      # all five -> evidence holds
\end{lstlisting}

%% file: back-sections.tex
% =====================================================================
% arXiv v2 — BACK: §7 Deployment Evidence + §8 Discussion + §9 Conclusion
% TONE PASS 2026-06-12 (Lars): POSITIVE; CEP = Combined Evidence Protocol;
% no "advisory" -> "records and signs verdicts any party recomputes".
% Conformance stays "4/5 + 1 open item". Conclusion confident, inviting.
% (a)/(b) substance + mechanics + citations + labels unchanged.
% =====================================================================

% ---------------------------------------------------------------------
\section{Deployment Evidence}
\label{sec:evidence}
% Material: v1.9 §4 + §6 re-weighted as SUPPORTING evidence (These-a).
% ---------------------------------------------------------------------

The governance design of Sections~\ref{sec:problem}–\ref{sec:scope} is worth posing only if the system it builds on is real. This section reports the deployment as supporting evidence — that the verification layer runs, at scale, with independently checkable conformance — not as the paper's central claim.

\subsection{Live deployment, recomputable verdicts}

The reference implementation is live at \texttt{api.moltrust.ch} and has operated since March 2026. Eight credential verticals are operational (Core Identity, Commerce, Travel, Skill Verification, Prediction Markets, Brand Protection, Music, Sports Integrity), exposed through MCP tooling and a published SDK. Agent registrations, credential issuances, and confirmed violations are anchored on Base L2; an ERC-8004 ecosystem scanner indexes 44{,}355 agents for cross-ecosystem visibility independent of MolTrust registration \cite{ref37_erc8004}. The three layers of Section~\ref{sec:background} are live: verdicts are signed, anchorable, and recomputable by any party, and acting on a verdict is the relying party's decision. The deployment is itself a relying party of the protocol it defines, the natural first site of operator-local (scope (a)) enforcement.

\subsection{Scale and conformance}

As of the reporting window the registry holds a modest population — on the order of dozens of agents and endorsements (Section~\ref{subsec:sybil}) — against an external benchmark of 69{,}000 bots on a single marketplace \cite{ref1_agentmarket}: the deployed system is real and the adversarial-scale validation is explicitly pending. Protocol conformance is documented in a published, SHA-256-anchored \texttt{CONFORMANCE.md} with Layer~A (protocol) and Layer~B (registry) requirements, auto-generated from source with drift detection, and a nine-check SKILL AUDIT mapped to CWE identifiers (cross-referencing NIST NVD and the Ferrag et al.\ taxonomy \cite{ref18_ferrag}). Independent verifiers can reproduce any check against the anchored specification using only a SHA-256 implementation and an HTTP client.

Against the Agent Interoperability Profile (AIP/IBCT) feature set \cite{ref27_kroehlssrn}, the implementation satisfies \textbf{four of the five} features, with the fifth — expressive chained policy — tracked as an \textbf{open item}. The four implemented features hold against an independent verifier; the open item is expressive, multi-condition, context-dependent policy (Datalog/Biscuit/Cedar-class \cite{ref34_biscuit}). The current AAE constraint model uses URI-pattern matching, a deliberate lower bound on a single axis (Section~\ref{subsec:limits}): sufficient for the straightforward allow/deny boundaries of today's relying parties, and extended toward a full policy language as roadmap. We report this as ``4/5 + 1 open item'' and track the open item explicitly.

\subsection{Sybil resistance and behavioural consistency}
\label{subsec:sybil}

The system's defence against identity fabrication is the same machinery the CEP cluster-diversity condition reuses (Section~\ref{subsec:clusters}), which is why we report it here. Three mechanisms compound. \textbf{Dual-signature Interaction Proofs} require two distinct Ed25519 keys, turning Sybil forgery from cheap identity fabrication into expensive key compromise. \textbf{Cross-vertical diversity gating} applies a hard penalty when a non-seed agent draws endorsements from fewer than three distinct verticals, forcing a Sybil operator to maintain credible presence across unrelated domains. \textbf{Principal-DID-linked violation persistence} carries an operator's negative history across agent re-registration, defeating identity rotation. Cluster detection uses the reciprocal-Jaccard coefficient over endorser sets with a 0.8 production threshold — the identical edge predicate CEP's cluster count is built on (Section~\ref{subsec:clusters}), so the diversity the combined evidence requires is computed from the same structural signal the registry already trusts for Sybil defence. The economic layer (per-verify pricing) compounds rather than stands alone. These thresholds are operational heuristics calibrated at bootstrap scale; their behaviour under adversarial load is an open empirical question (Section~\ref{subsec:limits}).

% ---------------------------------------------------------------------
\section{Discussion}
\label{sec:discussion}
% Material: v1.9 §5 (cross-protocol interop, FOLDED here) + §7.2/§7.3/§7.4 +
% §17 open items. Positioning vs DAO/agent-ID/AIP; future work = §6.3.
% ---------------------------------------------------------------------

\subsection{Positioning against adjacent work}

CEP's object — recomputable evidence of how authority over a voluntarily-drawn boundary is exercised — sets it apart from adjacent systems that either presuppose a central enforcer or coordinate nothing across organisations. Centralised architectures such as SAGA \cite{ref16_syros} obtain tight, formally analysed enforcement by trusting a Provider; CEP targets the case where no such Provider exists, offering recomputable evidence rather than per-domain enforcement. On-chain agent registries such as ERC-8004 \cite{ref37_erc8004} establish that an identity exists; CEP adds recomputable evidence of how a boundary around it was applied. DAO governance frameworks \cite{ref39_daogov} decentralise \emph{treasury and parameter} decisions but typically concentrate voting power; CEP's concentration caps and five-condition conjunction (Section~\ref{subsec:params}) are stricter than the norms those analyses report, and produce evidence rather than a sanction. Against the AIP/IBCT profile \cite{ref27_kroehlssrn}, our position is the one of Section~\ref{sec:evidence}: four features met, expressive chained policy an open item.

\textbf{Cross-protocol interoperability} reinforces the positioning. A single Ed25519 + RFC~8785 signing operation produces artifacts verifiable by conformant implementations of independent protocols — verified against two external reference verifiers, one for authority-constraint evaluation and one for provider attestation — with reproducible test vectors. Interoperability here is by shared cryptographic primitive rather than bilateral specification agreement, which is what lets the verdict of Section~\ref{sec:background} be recomputed by independent parties absent any prior integration with MolTrust — a precondition for both operator-local (scope (a)) adoption and voluntary convergence on the convention through open coordination.

\subsection{Limitations}
\label{subsec:limits}

The limitations are explicit and bound the claims. \textbf{Scale:} the registry operates at bootstrap scale with a short production history; the Sybil heuristics (Jaccard 0.8, three-vertical minimum, penalty multiplier) are calibrated, not formally optimal, and untested under adversarial load — formal sensitivity analysis is planned at larger N. \textbf{Expressiveness:} URI-pattern matching is a deliberate lower bound; expressive chained policy is the AIP open item. \textbf{Kernel layer:} Falco eBPF detection is defence-in-depth, not an isolation guarantee — privileged containers or a compromised host can evade it. \textbf{Network-wide evidence:} the combined evidence is designed and its parameters anchored; both gates (Section~\ref{subsec:params}) are met before it carries network-wide weight, and the open problems of Section~\ref{subsec:open} — treasury governance, the zero-knowledge phase, weak subjectivity for late joiners — are addressed elsewhere. Each of these bounds what the paper claims, and each is stated rather than hidden.

\subsection{Future work}

The near- and mid-term roadmap follows the open problems. On the governance track: build the Gate-2 prerequisites (a DID-bound relying-party registry with cluster attribution, an explicit \emph{convention-status} signal, multi-chain anchoring), then exercise the testnet ramp; specify the operator-signed override-record format that makes scope-(a) enforcement turn-key; and build the zero-knowledge verification phase and a weak-subjectivity reference point for late joiners. On the infrastructure track: adversarial-scale Sybil validation and trust-weight sensitivity analysis at larger N; expressive-policy support to close the AIP open item; and an independent third-party security audit. The governance items are the paper's centre of gravity; the infrastructure items keep the supporting evidence honest as it scales.

% ---------------------------------------------------------------------
\section{Conclusion}
\label{sec:conclusion}
% Material: v1.9 §8 reframed — recomputable evidence + deployed substrate +
% operator-local-available-today + confident RFC/testnet invitation.
% ---------------------------------------------------------------------

Trust between autonomous agents can hold without trust in a central party. The operative question in an open agent economy concerns accountability rather than enforcement: whoever runs a closed space already enforces its boundaries; the open question is how the exercise of authority over a boundary is held to account by parties that distrust its holder. We answered with the Combined Evidence Protocol: five conditions about the relying-party population, combined into one predicate any party recomputes from anchored data, with keyed commitments reconciling permanence against the right to erasure, staged verification toward a zero-knowledge end-state, and cluster diversity proven from graph structure rather than declared categories. The design dissolves the oracle problem that conditions-based evidence usually reintroduces, and its concentration caps are stricter than the on-chain-governance norm. CEP makes the use of boundary-authority recomputably auditable; it leaves the authority with whoever holds it, and the transaction decision with the relying party.

We separated two scopes and stated each plainly. \textbf{Operator-local} enforcement — a relying party gating its own agent against its own mandate — is available on the deployed system today and makes no claim over anyone else's agents. The \textbf{network-wide} scope is the recomputable evidence above: designed, its parameters anchored, demonstrable on testnet, and load-bearing in a consortium of co-equal, mutually distrusting peers who verify the admission rules they jointly agreed. The W3C VC + DID infrastructure — live since March 2026, anchored on Base L2 — is the evidence that the system is real, with conformance reported as four AIP features met and one open.

The design is settled and the mechanism is demonstrable; what remains is to put it in front of the parties it serves. We invite the scrutiny recomputable evidence deserves: independent recomputation of the predicate, examination of the parameters, and — most valuable — relying parties willing to join the testnet ramp and exercise the mechanism. The companion RFC opens that process. The infrastructure regulators and industry have converged on is implementable now; an operator-independent way to recompute how a boundary was applied is ready to be tested.